\PassOptionsToPackage{hyperfootnotes=false}{hyperref}
\documentclass[a4paper,fleqn]{cas-dc}
\usepackage[utf8]{inputenc}
\usepackage[authoryear]{natbib}
\usepackage{microtype}
\usepackage{url}
\usepackage{placeins}
\usepackage{needspace} \usepackage{tabularx}
\AtBeginDocument{\urlstyle{same}}
\ExplSyntaxOn
\bool_gset_true:N \g_stm_nologo_bool
\RenewDocumentCommand \emailauthor { m m }
  { \int_gincr:N \g_ead_int
    \seq_gput_right:Nn \g_stm_ead_seq { \texttt{#1}\space(#2) } }
\RenewDocumentEnvironment { Abstract } { o }
  { \group_begin:
    \IfNoValueF{#1}{\tex_gdef:D\abstractname{#1}}
    \parindent=0pt
    \box_if_empty:NTF\g_stm_key_box
      {\leftskip=.35\textwidth}
      {\dim_gset:Nn\l_tmpa_dim{\box_ht:N\g_stm_key_box}
       \dim_gadd:Nn\l_tmpa_dim{\box_dp:N\g_stm_key_box}
       \leftskip=.35\textwidth \hspace*{-.35\textwidth}
       \noindent\hbox_to_wd:nn{0pt}{\box\g_stm_key_box\hss}
       \skip_vertical:n{-\l_tmpa_dim}}
    \noindent\abstractname\par\skip_vertical:n{-4pt}
    \noindent\rule{.65\textwidth}{.2pt}\par\footnotesize
    \ignorespaces\everypar{\parindent=1.5em}
  }{\par\group_end:}
\ExplSyntaxOff

\makeatletter
\renewcommand{\bibinfo}[2]{#2\ifstrequal{#1}{pages}{\ifstrequal{#2}{1--35}{.\space}{}}{}\ifstrequal{#1}{journal}{\@ifnextchar,{\nobreak}{\space}}{}}
\makeatother
\newcommand{\pendingfigure}[3]{%
  \IfFileExists{#1}{\includegraphics[width=\linewidth]{#1}}{%
    \setlength{\fboxrule}{0.7pt}\setlength{\fboxsep}{0pt}%
    \fbox{\parbox[c][#2][c]{\dimexpr\linewidth-2\fboxrule-2\fboxsep\relax}{%
      \normalfont\centering
      \ttfamily\footnotesize [ARTWORK PENDING]\par\vspace{4pt}%
      \normalfont\itshape\footnotesize #3\par\vspace{4pt}%
      \normalfont\ttfamily\scriptsize\detokenize{#1}\par}}%
  }%
}
\newcommand{\D}[1]{#1}
\newcommand{\pending}[1]{}
\hypersetup{pdftitle={An LLM-powered Agent Framework for Heterogeneous Evacuation Behavior Modeling under a Moving Threat in a Public Plaza}}
\begin{document}
\let\WriteBookmarks\relax
\def\floatpagepagefraction{.8}
\def\textpagefraction{.1}
\shorttitle{LLM-powered Modeling of Evacuation under a Moving Threat}
\shortauthors{Ma, Yu, Tang and Li}
\title[mode=title]{\D{An LLM-powered Agent Framework for Heterogeneous Evacuation Behavior Modeling under a Moving Threat in a Public Plaza}}
\author[1]{Jian Ma}
\author[1]{Runxin Yu}
\author[1]{Tianyu Tang}
\author[2]{Xiaolian Li}
\affiliation[1]{organization={School of Transportation and Logistics, Southwest Jiaotong University},city={Chengdu},postcode={610031},country={China}}
\affiliation[2]{organization={Fujian Police College},city={Fuzhou},postcode={350007},country={China}}

\begin{abstract}
Modeling heterogeneous evacuation behavior under a moving threat is challenging because human perception, memory, and evidence evaluation resist representation through fixed rules. To overcome this limitation and capture these internal decision processes, we propose a novel agent-based framework powered by large language models (LLMs). It has three main components: (i) each pedestrian agent perceives a private symbolic ASCII view, mantains a Memory-based Knowledge Graph updated solely via individual observations, and makes decisions using prompts that incorporate a personality profile under a common sampling configuration; (ii) a decision-context compression coupled with a stateless memory model preserves trial-and-error experience across decision turns while omitting reasoning traces to keep the context compact, with the resulting memory record reconstructed post hoc as a knowledge graph; and (iii) a validation engine executes routes exclusively over observed terrain, effectively decoupling behavioral choice from physical feasibility. In a simulated public plaza scenario, we evaluated eight personality compositions in eight paired randomized blocks. The results reveal a strong association between usable-exit knowledge and evacuation outcomes: agents with such knowledge evacuated in 89.5\% of cases, compared to only 1.05\% of those lacking it. Furthermore, personality compositions differed in their evaluation of threat evidence. Under remembered evidence, the proportion of danger assessments varied by 0.265 across personality compositions, while the mean share of high-urgency, low-directness decisions varied approximately three-fold, from 11.41\% to 34.33\%. By contrast, responses converged after direct threat sightings: 99.6\% of the assessments classified the situation as dangerous, as instructed. Movement was selected in 99.8\% of decisions. Overall, the results indicate that access to information was strongly associated with evacuation outcomes, while spatial geometry, information, and affect were jointly associated with evacuation time. The proposed framework provides an auditable approach to generating endogenous behavioral heterogeneity through persona-conditioned LLM agents in crowd-evacuation simulations.
\end{abstract}


\maketitle
\hypersetup{
pdfauthor={Ma, Yu, Tang and Li},
pdfcreator={LaTeX; Elsevier cas-dc 2.4},
pdfsubject={LLM-powered evacuation modeling under a moving threat},
pdfkeywords={Crowd Evacuation Simulation, Large Language Models, Symbolic Spatial Representation, Agent Memory, Personality Traits, Public Space Security}}

\section{Introduction}\label{sec:1}

Public plazas, temple forecourts, transport interchanges, and other open civic spaces bring large numbers of people into layouts designed for circulation. When a threat emerges within such a space, people must rapidly determine what is happening, whether they are at risk, where the exits are, and how those nearby are likely to respond. Pedestrian safety in these settings has therefore attracted sustained attention across disciplines. Computer-science research detects congestion from motion signatures in trajectory data \citep{khansd2019}, while emergency-management research develops actionable evacuation plans for dense facilities \citep{guok2023}. A moving threat intensifies the evacuation problem: the hazard follows a dynamic path, information reaches different individuals incompletely and at different times, and route choice depends on spatial cognition, risk perception, and geometry of the scene. Consequently, mathematical and computational models have sought to capture how population heterogeneity shapes evacuation behavior.

Existing evacuation models operate at three scales: macroscopic, mesoscopic, and microscopic. Macroscopic models represent a crowd as a homogeneous flow described by density, velocity, and flow rate; common formulations include queuing, regression, and path-selection models \citep{kaurn2022}. Li et al., for example, developed a numerical scheme based on the Hughes continuum model that relates simulated density to physical compression forces \citep{lix2018}. Individual heterogeneity is difficult to represent at this scale \citep{kouskoulisg2018}. Mesoscopic models occupy the intermediate scale: rather than tracking each evacuee, they describe the crowd through a statistical distribution over individual states such as position, velocity, and walking direction, typically in gas-kinetic or kinetic-theory form \citep{hoogendoorns2000,bellomon2011}. Agnelli et al., for example, applied kinetic theory to crowd evacuation from bounded domains \citep{agnellijp2015} and later extended the approach to disease contagion during evacuation \citep{agnellijp2023}. Heterogeneity at this scale is typically represented through a small number of activity variables or state classes, so individual perception and decision processes remain outside the model. Microscopic models instead represent evacuees individually and can therefore capture individual attributes and local interactions. Social-force models \citep{helbingd2000}, cellular automata \citep{kirchnera2003}, and hybrids of the two are standard examples. Chen et al. developed social-force formulations considering the spatial features of tread depth and riser height in staircases to explore vertical evacuation of buildings \citep{chenj2018}. These formulations resolve mechanical interactions within physical space, but cognitive processes, perceived information, and behavioral preferences generally remain external to the model. Subjective trade-offs among threat exposure, congestion, and travel distance must therefore be encoded as rules.

Pedestrian microsimulation generates behavior from predefined physical or behavioral rules affected by environmental features and personal behavioral preferences. Recent work has extended this rule-based foundation in several directions. Simulation-based optimization has been coupled with real-time evacuation guidance under changing access conditions and crowd distributions \citep{zhangb2024}; visibility-dependent interaction models reproduce detours, wall following, and herding under reduced visibility \citep{tanj2025}; and machine-learning models infer microscopic pedestrian motion from local interaction data \citep{jiangn2025}. Multi-agent studies have also compared situational-awareness and routing strategies under dynamic hazards \citep{serranoa2026}, as well as heterogeneous exit-choice strategies that balance efficiency, safety, and fairness \citep{tongy2026}. Collectively, these studies make information, perception, and behavioral heterogeneity increasingly explicit, but decisions remain encoded in optimization objectives, learned functions, or prescribed strategy sets. Agent-based modeling has consequently attracted interest as a means of representing human-like behavior and the cognitive processes underlying individual decisions \citep{gaoc2024}. Psychological structure has been introduced into agent-based evacuation models through Gibson's theory of affordances for pedestrian--environment interaction \citep{hassanpours2021}, trained leaders who redistribute crowd route choices \citep{pelechanon2006}, and heterogeneous self-driven forces \citep{wuw2022}. Nevertheless, these formulations rely on weighted functions and manually specified rules that define the population's behavioral repertoire in advance.

Large language models (LLMs) offer a way to condition agent decisions on descriptions of the environment, individual attributes, and remembered experience. Research on language processing has identified shared computational principles between human neural responses and language models, including context-dependent prediction and representations of word meaning \citep{goldstein2022shared}. More recently, a language model fine-tuned on human behavioral data predicted choices across a range of cognitive tasks \citep{binz2025cognition}, supporting interest in language models as tools for studying human decision-making.  Interactive agents extend this potential to decisions made through repeated encounters with an environment. In Minecraft, JARVIS-1 combines planning, action execution, and memory of previous experiences to complete tasks requiring multiple stages of action \citep{wang2025jarvis}. For evacuation modeling, this division of functions offers a way to connect immediate responses to danger with continuing goals and revisions to an escape plan.  Recent studies have applied LLM-powered agents to evacuation and wayfinding. Dang et al. combined LLMs with a cellular-automaton environment to examine how agent backgrounds and communication relate to fire-evacuation outcomes \citep{dangp2025b}. In a separate wayfinding study, they compared spatial-memory conditions and included human participants as a reference \citep{dangp2025a}. Yang et al. incorporated personality traits, environmental observations, and action histories into an LLM-powered evacuation framework \citep{yangs2026}. Explicit spatial and knowledge representations provide additional support for such decisions: neuro-symbolic evacuation models combine learned reasoning with spatial structure \citep{bahamida2024}, while Tao et al. coupled a fire-incident knowledge graph constructed through entity and relation extraction with vision-language-model reasoning for emergency decision support \citep{taoj2026}.  Building on this work, the present study asks how changing information is interpreted, retained, and translated into successive evacuation decisions under a moving threat. A central issue is how a pedestrian's understanding of danger and available exits develops as new evidence arrives. Connecting these changes to route choice and movement provides a basis for investigating the processes through which evacuation behavior develops.

Such a framework should meet at least the following four requirements. First, the environment must be presented in a form the model can interpret without devoting excessive capacity to decoding geometry. Second, population heterogeneity must arise within the agents, because demographic, physical, and psychological differences produce distinct evacuation behaviors \citep{hux2023,lis2023,mandarm2016}. Knowledge-guided hierarchical multi-agent reinforcement learning represents this heterogeneity through mobility attributes, sensitivity parameters, and decision mechanisms \citep{sunl2026}, while other work encodes personality traits as agent attributes \citep{wangh2021,raetanoj2026}. Third, each agent requires persistent, inspectable memory so that a decision can be traced to what the agent knew when it was made. Fourth, information must pass from person to person to reflect the uneven diffusion of knowledge through a crowd.

This study develops an LLM-powered agent framework for investigating how individual experience enters evacuation decisions in a public plaza under a moving threat. Its contributions are to represent decisions using private observations, personality profiles, and retained experience; to reconstruct changes in memory alongside successive choices; and to connect these records to physically executed movement. Together, these elements support the analysis of how pedestrians interpret danger, maintain an escape goal, and reconsider a route as their understanding of the situation changes. Section~\ref{sec:2} describes the framework, Section~\ref{sec:3} presents the experimental design, Section~\ref{sec:4} reports the results, Section~\ref{sec:5} discusses their implications and limitations, and Section~\ref{sec:6} concludes the paper.

\section{Methodology}\label{sec:2}

\begin{figure*}[pos=tp]
\centering
\includegraphics[width=\linewidth]{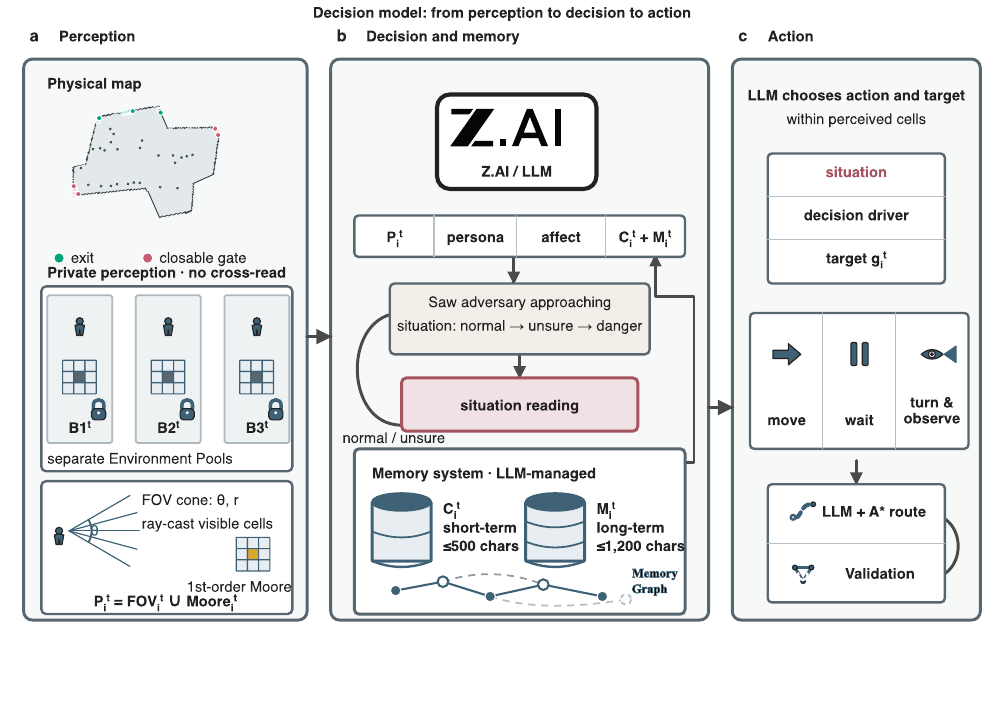}
\caption{Perception, decision, memory, and physical execution in the LLM-powered framework. (a) Local observations update each pedestrian's private known map $B_i^t$. (b) The decision model combines the current view $P_i^t$, personality, affect, short-term context $C_i^t$, and long-term memory $M_i^t$. (c) The model selects an action and target $g_i^t$; the engine constructs and checks routes through observed, passable cells. Memory revisions are reconstructed as knowledge graphs after the run.}
\label{fig:framework}
\end{figure*}

The framework separates local perception, decision-making, memory updating, and physical execution. Each pedestrian maintains a private known map $B_i^t$ containing observed terrain. The decision model receives the current symbolic view $P_i^t$, the persona description, the engine-computed affective state, short-term context $C_i^t$, and long-term textual memory $M_i^t$. It selects an action and either an offered route or a perceived target. The engine generates and validates routes using the pedestrian's known map, then applies the physical movement rules. After each run, memory revisions and decision records are reconstructed as graphs for analysis. Figure~\ref{fig:framework} distinguishes the information used during simulation from the records constructed afterward.

\subsection{Personality Modeling and Prompt Specialization}\label{sec:2.1}


The framework represents stable behavioral tendencies using the Five-Factor Model (FFM): openness to experience (O), conscientiousness (C), extraversion (E), agreeableness (A), and neuroticism (N). The FFM provides a compact description of individual differences \citep{raetanoj2026}. Evacuation models have used OCEAN traits to relate personality to psychological stress and desired speed \citep{wangh2021}, while separate work has examined aggressiveness and competition in pedestrian movement \citep{hux2023}. In the present framework, the five trait scores form part of each agent's persistent persona description.

\subsubsection{Personality Configuration}\label{sec:2.1.1}

With the five-domain FFM profile, each evacuee's personality can be represented as an agent attribute expressed as T-scores. The score is embedded for each agent in the system prompt received by the decision model on every call. Each profile assigns a T-score (mean 50, standard deviation 10) to openness, conscientiousness, extraversion, agreeableness, and neuroticism. Scores are mapped to five interpretive bands: very low (below 35), low (35--44), average (45--55), high (56--65), and very high (above 65). For each domain, the prompt includes the numeric score, its band, and a behavioral anchor framed in the context of an ordinary outing. 

For example, the high-conscientiousness anchor describes a person who adheres to a plan and follows a chosen route, whereas the high-neuroticism anchor describes someone who hesitates and revisits choices when a situation is unclear. Domains outside the average band also include salient facets that link non-average scores to specific tendencies. Table~\ref{tab:persona} presents the five profile templates used to instantiate this representation. Population size and the number of pedestrians assigned to each profile are specified separately for each experimental scenario.

\begin{table}[pos=htbp]
\caption{FFM T-scores of the five personality archetypes.}
\label{tab:persona}
\centering
\small
\setlength{\tabcolsep}{4pt}
\renewcommand{\arraystretch}{1.16}
\begin{tabular}{@{}>{\raggedright\arraybackslash}p{.34\linewidth}ccccc@{}}
\toprule
\textbf{Archetype} & \textbf{O} & \textbf{C} & \textbf{E} & \textbf{A} & \textbf{N} \\
\midrule
Average & 50 & 50 & 50 & 50 & 50 \\
Calm-independent & 50 & 55 & 40 & 45 & 30 \\
Rule-following & 40 & 70 & 45 & 60 & 50 \\
Anxious-conformist & 40 & 45 & 62 & 60 & 72 \\
Bold-solitary & 68 & 35 & 35 & 40 & 32 \\
\bottomrule
\end{tabular}
\end{table}

The five personality profiles use the fixed T-scores listed in Table~\ref{tab:persona}. These scores determine the interpretive bands, salient facets, and behavioral anchors inserted into each agent's identity section. The neuroticism score also determines the sensitivity and recovery coefficients in Eq.~\eqref{eq:neuroticism}. The composition experiment changes the numbers of pedestrians assigned to each profile while keeping the profile scores fixed.

\subsubsection{Behavioral Differentiation through Persona-Conditioned Decisions}\label{sec:2.1.2}


Rule-based evacuation models usually represent individual differences through parameter values or prescribed behavioral rules. In the present framework, by contrast, behavioral heterogeneity is introduced through a stable persona profile embedded in each agent's system prompt. The persona conditions the decision but does not prescribe a fixed action; pedestrians presented with the same environmental evidence may therefore still make different choices. The language model interprets this profile together with current perception, affect, and remembered experience, and this interpretation is the core mechanism through which different agents express their heterogeneity.

Agents accumulate different observations and memory texts as their trajectories diverge. These changing inputs can accompany differences in later decisions. The graphs reconstructed after a run describe this recorded history. As a result, individual differences in personality become progressively more pronounced. This phenomenon emerges from the long-term effects of LLM-powered self-maintenance, through which perceptual, decision-making, and other behavioral information are progressively integrated over time, ultimately giving rise to an increasingly coherent and realistic perception–decision–action framework for each individual.

\begin{figure*}[pos=tp]
\centering
\includegraphics[width=\linewidth]{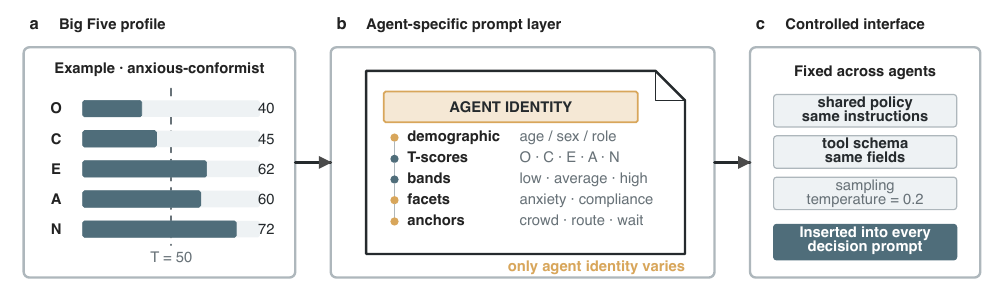}
\caption{Personality-to-prompt encoding in the controlled decision interface. (a) An anxious-conformist profile illustrates the five FFM T-scores and their position relative to $T=50$. (b) The agent-specific identity section converts demographics and scores into bands, salient facets, and behavioral anchors. (c) The shared policy, tool schema, and \texttt{temperature}=0.2 remain fixed across agents, and the resulting identity section is inserted into every decision prompt.}
\label{fig:persona}
\end{figure*}

Fig.~\ref{fig:persona} summarizes how the five trait scores are translated into agent-specific descriptions while the general decision instructions and sampling settings remain fixed across agents.

For each agent, its system prompt is composed of a shared policy section and an agent-specific identity section. The shared policy is identical for all the agents and consists of seven instruction blocks, summarized in Table~\ref{tab:policyblocks}, while the identity section varies depending on the agent's demographic type and FFM profile. For each decision the agent made, the system prompt load  varies because it contains the agent's updated current perception, engine-computed affect, and retained memory. This separation makes the sources of prompt variation explicit, but it does not isolate the causal effect of persona text because neuroticism also affects the engine-maintained fear state described in Section~\ref{sec:2.3.5}.

\begin{table}[pos=htbp]
\caption{The seven instruction blocks of the shared policy section.}
\label{tab:policyblocks}
\centering
\small
\setlength{\tabcolsep}{4pt}
\renewcommand{\arraystretch}{1.16}
\begin{tabular}{@{}>{\raggedright\arraybackslash}p{.22\linewidth}>{\raggedright\arraybackslash}p{.70\linewidth}@{}}
\toprule
\textbf{Block} & \textbf{Content} \\
\midrule
Identity & An ordinary pedestrian acting on local perception, its own known map, memory, affect, and messages received \\
Map & The fixed north-up orientation, the \texttt{@}-relative offset convention, and the glyph table of Table~\ref{tab:glyphs} \\
Judgment & Target choice belongs to the agent, route legality to the engine; the evidence clause; the default target-distance band \\
Intent & \texttt{intent} and \texttt{urgency} as a standing goal that persists until the agent changes it \\
Route options & Fields of the engine-computed routes (length, familiarity, crowding, clearance from \texttt{A} or \texttt{D}, reach) and the rule that the shortest route need not be chosen \\
Past experience & Use of the decision-context capsule and of the engine's feedback on the previous response \\
Output & Validation and JSON action guideline \\
\bottomrule
\end{tabular}
\end{table}
The identity section of the system prompt describes each agent's identity and personality and remains fixed throughout the simulation. It includes the agent's identifier and demographic type, the five personality domain scores and their bands, the salient facets of domains outside the average range, and a brief description of a behavioral tendency for each domain. For example, the high-neuroticism description characterizes a preference for nearby targets and more frequent changes in movement direction under threat. These descriptions provide a personality context that the decision model considers together with local observations, memory, and affect when choosing an action.

At $t=0$, each pedestrian selects an initial route and enters the first recorded frame already moving (Section~\ref{sec:2.3.1}). The engine accepts a target when it lies on an observed, passable cell and can be reached through observed cells (Section~\ref{sec:2.2.3}).

During the simulation, an agent's move is affected by its surrounding environmental situation, especially the situation within its field of view. A direct sighting of the threat or an injured person provides unambiguous evidence that the current situation is dangerous. In the symbolic view described in Section~\ref{sec:2.2.3}, the glyph \texttt{A} marks an armed person rushing toward the crowd, and \texttt{D} marks a person lying motionless on the ground (Table~\ref{tab:glyphs}). For example, an agent may interpret \texttt{A} as ``I saw someone holding a knife with blood on it''. The decision instruction states that directly seeing \texttt{A} or \texttt{D} is strong evidence of danger. The label itself, however, is produced by the language model rather than written by the engine: the model must recognize the glyph in its own ASCII view and return \texttt{danger} in the \texttt{situation} field of its JSON output, which records its appraisal as \texttt{normal}, \texttt{unsure}, or \texttt{danger} (Section~\ref{sec:2.3.5}). The LLM then selects the agent's next action by considering its personality description, affective state, remembered experience, and visible surroundings.  Within each decision turn, the LLM interprets the symbolic observations, reports its appraisal of the situation, and selects an action. As new evidence becomes available, it can reconsider its standing goal and choose a different route or target in light of remembered experience. This connects cue interpretation, situation appraisal, and action selection within an LLM-powered decision process (Section~\ref{sec:2.3.2}).

Person-to-person communication is represented by shouting. Once a pedestrian has chosen to escape, it shouts while moving. The engine calculates the distance between the speaker and nearby pedestrians and adds the signal to the decision context of listeners within hearing range (Section~\ref{sec:2.3.7}). The signal conveys the direction and approximate distance of the shout. In everyday terms, the listener hears someone shouting nearby. Anonymous shouting is presented as uncertain evidence, which the listener's LLM considers alongside its observations, memory, personality, and affect. The model returns an updated appraisal of normal, unsure, or danger, and the engine records this as the pedestrian's current situation. This communication mechanism separates warning reception, interpretation, and the choice of a response \citep{lindell2012padm}.

All agents share the same sampling settings, so variation across personalities arises from trait scores, verbal anchors, affect, memory, and current evidence.

\subsection{ASCII-Based Local Perception and Spatial Representation}\label{sec:2.2}

Agents in the proposed framework receive their surroundings through a symbolic representation. By making decision-relevant information explicit at the cell level, this interface allows each move to be related to what the agent could observe \citep{ahissare2025,bahamida2024}. This section defines the environment layers that generate the view, the geometry that determines which cells enter it, and the encoding that converts those cells into prompt text, as shown in Fig.~\ref{fig:perception}.

\subsubsection{Layered Environment Representation}\label{sec:2.2.1}

The environment comprises four layers with distinct functions. The 1st layer is the raw layer, a raster of 0.4~m cells that records the fixed geometry of the space. The 2nd layer is the semantic layer, which records openings, gates, fixed obstacles, and spawn cells. The 3rd layer is the effective layer, which combines the raw and semantic layers with the current gate states and serves as the physical authority for collision and availability. The 4th layer is the belief layer, which is private to each agent and contains only terrain that the agent has personally observed.

The reason why we separate the belief layer from the effective layer is that it establishes the framework's information boundary. Physical resolution uses the effective layer, whereas target selection and local route generation use the pedestrian's private belief layer. The model receives only the privately observed and explored map, and the memory connects the selected target exclusively through observed cells.

\subsubsection{Field of View and the perception map}\label{sec:2.2.2}

Visibility is computed by enumerating every raster cell whose center lies within the sight radius and testing it against the view cone and a line-of-sight trace. Let $\mathcal{C}$ denote the static set of raster cells and let $G^t$ denote the effective physical-grid state at time $t$. For pedestrian $i$, with position $\mathbf{x}_i^t$, heading $\mathbf{h}_i^t$, sight radius $R_i$, and cone width $\varphi_i$, the visible set of the environment is
\begin{equation}
\begin{split}
F_i^t=\bigl\{\,q\in\mathcal{C}\ \big|\ &
\lVert\mathbf{x}_q-\mathbf{x}_i^t\rVert\le R_i,\\
&\angle\bigl(\mathbf{x}_q-\mathbf{x}_i^t,\mathbf{h}_i^t\bigr)
\le \varphi_i/2,\\
&\mathrm{LOS}(\mathbf{x}_i^t,\mathbf{x}_q;G^t)\,\bigr\}.
\end{split}
\label{eq:fov}
\end{equation}
The perception map also includes the first-order Moore neighborhood $\mathcal{N}_i^t$, representing immediate awareness of the eight cells surrounding pedestrian $i$ at time $t$, including cells behind the pedestrian:
\begin{equation}
P_i^t=F_i^t\cup\mathcal{N}_i^t.
\label{eq:perceived}
\end{equation}

\begin{figure*}[pos=tp]
\centering
\includegraphics[width=\linewidth]{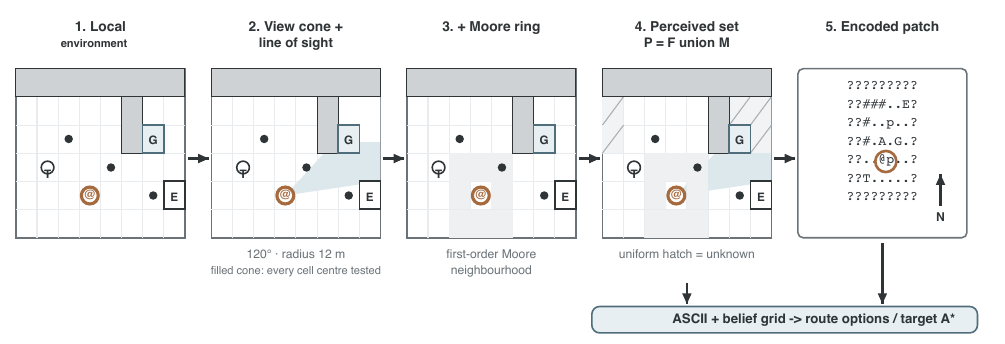}
\caption{From geometry to action. The filled view cone and first-order Moore neighborhood form the perception map, unknown cells appear as \texttt{?}, and the private Memory-based Knowledge Graph supports engine-generated route options as well as A* validation of a model-selected target.}
\label{fig:perception}
\end{figure*}

At initialization, each pedestrian also observes the cells within a Chebyshev radius of 3, subject to line of sight. These initial terrain observations are stored in the memory section and serve as  one part of system prompts before the first decision and persist after ordinary directional perception begins. A wall cell remains visible while occluding cells behind it, and every cell outside $P_i^t$ appears as \texttt{?}. Moore-neighborhood cells are integrated into the same equal-width ASCII matrix, as shown in Fig.~\ref{fig:perception}. At each simulation timestep, $P_i^t$ is written to the private Memory-based Knowledge Graph together with its observation time. The resulting Memory-based Knowledge Graph supplies route options and supports validation of model-selected targets.

\subsubsection{Symbolic Encoding of the Perceived Field}\label{sec:2.2.3}

The perception region aligns with the orientation of the agent's facing direction. The terrain from the effective layer is written first, after which entities are overlaid: other pedestrians, people on the ground, the anomalously behaving person, and the focal agent. Table~\ref{tab:glyphs} defines the vocabulary. Every cell in the bounding box but outside the perception map is written as \texttt{?}, making the boundary between observed and unknown space explicit.

\begin{table}[pos=htbp]
\caption{Character vocabulary of the local symbolic view.}
\label{tab:glyphs}
\centering
\small
\setlength{\tabcolsep}{5pt}
\renewcommand{\arraystretch}{1.16}
\begin{tabular}{@{}c>{\raggedright\arraybackslash}p{.72\linewidth}@{}}
\toprule
\textbf{Symbol} & \textbf{Meaning} \\
\midrule
\texttt{?} & Outside the perception map or occluded \\
\texttt{.} & Walkable ground \\
\texttt{\#} & Wall or outside the building \\
\texttt{T} & Tree or fixed obstacle \\
\texttt{E} & Exit \\
\texttt{G} & Gate, open \\
\texttt{g} & Gate, closing \\
\texttt{X} & Gate, closed \\
\texttt{@} & The agent itself \\
\texttt{p} & Another pedestrian \\
\texttt{D} & A person lying motionless on the ground \\
\texttt{A} & An abnormal person rushing toward the crowd \\
\bottomrule
\end{tabular}
\end{table}

The decision model returns an action, a situation assessment, a standing intent, and either an engine-provided route identifier or a target offset. The engine checks the selected destination against the saved decision-time observation, constructs an eight-neighbor A* route on the pedestrian's private Memory-based Knowledge Graph, and rejects unknown, blocked, or disconnected targets. Only validated belief-grid cells are executed. 

\subsubsection{Properties of the Two-Dimensional Symbolic Field}\label{sec:2.2.4}

A character grid encodes position and attribute in a single token. A symbol's location in the matrix specifies where an object is relative to the agent, while the symbol itself specifies the object's identity. Thus, reading \texttt{A} two rows above and one column to the right conveys both properties simultaneously. Throughout a run, textual adjacency preserves neighborhood relationships in the two-dimensional field.

The representation is also preserved under reorientation. When a scene is rotated, interpreting a coordinate list requires the spatial relationships to be re-derived arithmetically. By contrast, the character grid rotates with the scene while preserving layout and adjacency, so the reading procedure remains unchanged. This property allows the benchmark below to separate spatial competence from arithmetic competence.

Spatial-representation validation was retained as a supplementary experiment using frozen snapshots to compare four representations: no map, structured cells, lean ASCII, and ASCII with a decoding protocol. In the main evacuation experiment, the implemented equal-width ASCII representation remained fixed while the behavioral mechanisms were tested.

\subsection{LLM-powered Agent Modeling Framework}\label{sec:2.3}

Once the surroundings have been perceived, agents with different behavioral characteristics must decide how to act. Their information is partial, uncertain, and dynamic, as is typical in evacuation settings, where environmental cues, risk appraisal, and nearby behavior jointly inform the next move. Linguistic instead of purely numerical representation reflects how such situations are described and interpreted \citep{xues2023}. Language models, in turn, provide a mechanism for context-dependent decisions that can incorporate heterogeneous behavioral characteristics \citep{gaoc2024,yangs2026}. This section specifies the decision loop, supporting memory architecture, affective state, and algorithmic adversary.

\subsubsection{Perception--Decision--Action Loop}\label{sec:2.3.1}

Before the simulation starts, the decision model receives each pedestrian's initial observation and selects either an offered route or a legal target. The engine validates the selection and installs the resulting route, so each pedestrian enters the first recorded frame already moving along a model-selected route.

Each subsequent turn follows a fixed sequence. Memory revisions submitted during the preceding turn become due first. If a revision remains unfinished, the simulation waits at the current boundary before constructing any new decision observation. Fields of view are then recomputed and local evidence registered. Every active pedestrian carries a current route and may also have one pending decision request. A decision submitted at simulated time $t$ stores the pedestrian's observation and memory snapshot and is scheduled for the first boundary at or after $t+1$. During this interval, the pedestrian may execute one step along an existing legal route. An early response remains buffered until the due boundary; a late response pauses pedestrians, the adversary, gates, affect, collision recovery, trajectory recording, and the simulation clock until computation finishes. The validated decision is then applied at the same boundary before movement.

When a pedestrian first sees an armed person (\texttt{A}) or someone lying motionless (\texttt{D}), it reassesses the situation using the new observation. If a decision based on an earlier view is still being generated, that assessment is replaced by one that includes the newly observed evidence. While the threat remains visible, another assessment can begin once at least one simulated second has passed since the previous decision took effect. Each assessment uses the observation recorded at its start, and its decision takes effect one simulated second later. Subsequent changes in the scene are considered in the next assessment.

During pending computation, the pedestrian continues along an existing legal route and waits if that route is exhausted. After pedestrian decisions are applied, the adversary plans its movement, and all entities propose at most one move for concurrent resolution. Fatalities, opening capacity, evacuation, decision traces, metrics, and the rendered frame are then evaluated from the committed state.

\subsubsection{Dual-Model Architecture and Decision Layers}\label{sec:2.3.2}

Each agent is served by two models with distinct responsibilities. The decision model reads the symbolic view, compact working context, long-term memory, and affect; reports an intention, urgency, action, and situation assessment; and selects an offered route or perceived target. The memory model reads the agent's current notes together with its observations, decisions, and realized movement, then rewrites a first-person account. Separating these functions makes decision behavior and memory revision independently measurable \citep{dangp2025a}.

Within a single call, the decision model operates at three levels. At the highest layer, it states a standing goal: \texttt{intent} identifies what the agent is trying to do, while \texttt{urgency} indicates how strongly it judges that immediate action is required. Both values persist until the agent revises them. The previous values are returned to the agent on its next turn, so an intention continues by default and any change is explicitly recorded. At the middle layer, the agent selects a route. The engine constructs up to four walkable route segments on the pedestrian's private Memory-based Knowledge Graph, anchored to known openings or frontier cells at the edge of explored space. It also adds a corridor-distinct alternative to the nearest anchor, providing another path even when only one destination is known. Each segment is annotated with observable attributes: length in cells, the fraction already traversed, the number of visible people along it, the closest approach to any currently visible \texttt{A} or \texttt{D}, and whether the 20-cell horizon reaches the anchor or only advances toward it. Personality and affect can therefore influence whether the agent prioritizes familiarity, crowding, or clearance from a visible threat; the prompt explicitly gives no priority to the shortest option. If none of the offered routes is suitable, the agent selects a \texttt{target\_offset} and, optionally, one intermediate waypoint. At the lowest layer, the engine commits at most 20 cells of the selected route before requesting another decision, preventing a single choice from determining an entire long route.

\subsubsection{Context Compression and Experience Retention}\label{sec:2.3.3}

Each request is constructed from a fresh, self-contained context. Throughout the text and figures, $C_i^t$ denotes the decision-context window: per-agent short-term memory that carries compressed operational context between nearby turns. The symbol $M_i^t$ denotes per-agent long-term first-person memory maintained across revisions. The decision and memory models use these two written channels, respectively, to carry experience between turns.

The short-term memory is a rolling operational context. Its inputs are the previous content, the current observation, and the engine's feedback on the previous response, which states whether it was accepted and, if not, why it was rejected, so that the next decision can correct the error. The compression rule retains the current intention, strategies confirmed against the environment, tool corrections, and unresolved questions. It removes reasoning traces, restatements, expired map details, and relative coordinates invalidated by movement. Reasoning traces are omitted by design because they would not fit within the compression length limit; the capsule keeps the outcome of each attempt. A valid response replaces the memory prompts; a rejected response preserves the previous version.

The second model writes the long-term memory $M_i^t$ from the pedestrian's current memory text and a summary of the committed turn. The summary includes the number of visible people; identifiers of visible adversaries and bodies; signage and opening states within the perception map; affect and witnessed events; and the decision, including its target, action, and situation assessment. The model returns a first-person account within a 1{,}200-character limit, retaining current facts while revising or removing expired ones. Because facts persist only through rewriting, omission constitutes forgetting. When an adversary leaves the perceptual set, the pedestrian retains only the cell in which it last observed that adversary and the simulated time of the observation. Later requests label this record explicitly as stale evidence and report the elapsed time, with bearing and distance recomputed from the pedestrian's current position to the remembered cell. The record is created only while the adversary lies within that pedestrian's perceptual set and is replaced only by a subsequent direct sighting. Gaze and body direction are represented by a single heading, so a pedestrian cannot look behind itself without turning, and any committed step overwrites its previous orientation.

The two limits reflect different retention horizons. Operational context concerns the current situation and becomes stale within a few turns, so 500 characters are sufficient for information that survives one rewrite. Experience accumulates throughout a run, giving the memory model 1{,}200 characters within which to determine what the pedestrian retains.
\subsubsection{Graph-Structured Memory}\label{sec:2.3.4}

The memory model is stateless. Each update carries a complete context for one agent: its current memory text and the newly committed turn. Neither conversation history nor content is shared between agents. Revisions are written as prose, allowing the model to retain, modify, or omit remembered facts in the same format as the record. Facts absent from the current revision become forgotten nodes; forgetting is therefore endogenous to the agent and observable through logs that record what disappeared and when.

\begin{figure*}[pos=tp]
\centering
\includegraphics[width=\linewidth]{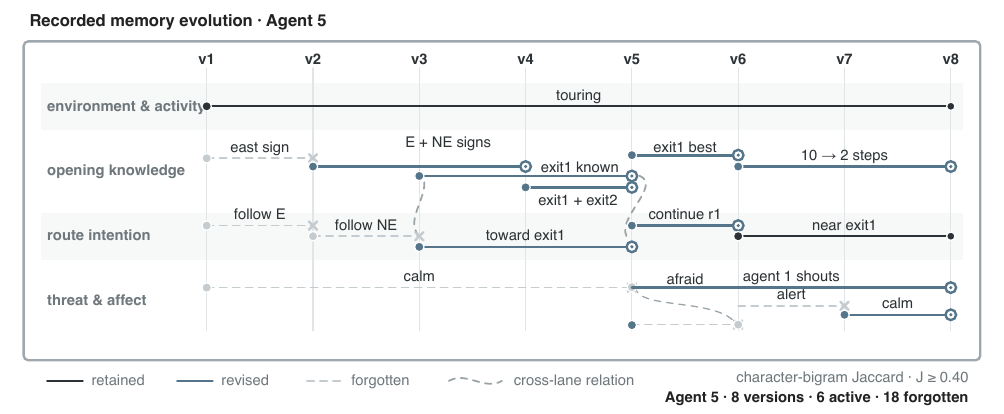}
\caption{Recorded evolution of Agent 5's long-term first-person memory in the completed GLM-5.3-flash reference run. Facts are arranged by version ($v_1$--$v_8$) and semantic track. Solid dark lineages were retained, blue nodes mark revisions, and pale dashed lineages terminate facts later omitted from memory. Dashed cross-track edges show recorded semantic associations, such as opening knowledge informing route intention and a heard shout accompanying an affect update. Reworded facts are matched by character-bigram Jaccard overlap at $J\geq0.40$. The final record contains six active and 18 forgotten facts; graph construction occurs after the simulation.}
\label{fig:memgraph}
\end{figure*}

The completed record is assembled into a versioned knowledge graph. Each revision is divided at sentence boundaries into individual remembered facts, each represented as a node. Facts are matched across versions using character-bigram Jaccard overlap,
\begin{equation}
J(a,b)=\frac{\lvert B(a)\cap B(b)\rvert}{\lvert B(a)\cup B(b)\rvert},
\label{eq:jaccard}
\end{equation}
where $B(\cdot)$ denotes the set of adjacent character pairs. This measure operates directly on the text and clearly separates the observed score ranges: rewritten versions of the same fact score 0.43--0.60, whereas distinct facts score 0.00--0.11. A threshold of 0.40 therefore assigns reworded facts to their existing nodes. A retained fact continues as one lineage across versions, a revised fact starts a new version within its lineage, and a fact omitted from a revision ends its lineage. Edges between lineages record semantic associations among opening knowledge, route intention, communicated evidence, and affect. Fig.~\ref{fig:memgraph} illustrates this structure for one recorded agent.

The knowledge graph is reconstructed after the simulation from the complete memory record. It connects changes in remembered information with recorded decisions, supporting analysis of how pedestrians revise their understanding and evacuation plans \citep{taoj2026}.

\subsubsection{Affective State}\label{sec:2.3.5}

The validation mechanism manages a fear value $f_i(t)\in[0,1]$ for each pedestrian $i$. Four categorical levels are used for reporting: calm below 0.20, alert below 0.45, afraid below 0.75, and high fear at or above 0.75. Stimuli arise from direct sightings, proximity to the threat, observed casualties, and nearby pedestrians fleeing. Let $z_{ik}(t)$ denote the intensity of stimulus type $k$ experienced by pedestrian $i$, and let $z_i^\ast(t)=\max_k z_{ik}(t)$ denote the strongest active stimulus. Fear is updated as
\begin{equation}
f_i(t+\Delta t)=
\begin{cases}
\max\!\left(f_i(t),\,\min\!\left(1,\kappa_i z_i^\ast(t)\right)\right),
& z_i^\ast(t)>0,\\[4pt]
\max\!\left(0,f_i(t)-\rho_i\Delta t\right),
& z_i^\ast(t)=0,
\end{cases}
\label{eq:affect}
\end{equation}
where the sensitivity multiplier $\kappa_i$ and recovery rate $\rho_i$ are
\begin{equation}
\begin{aligned}
\kappa_i&=\operatorname{clip}_{[0.65,1.45]}
\!\left(1+\frac{T_{iN}-50}{100}\right),\\[2pt]
\rho_i&=\max\!\left(0.01,\ 0.10-\frac{T_{iN}-50}{1000}\right).
\end{aligned}
\label{eq:neuroticism}
\end{equation}
A larger $T_{iN}$ therefore produces a stronger rise in fear and a slower recovery. The current fear value and active stimuli are supplied to the decision model, creating an explicit second pathway from the personality representation to later decisions.

The framework represents emotion using two quantities with distinct origins. The scalar $f$ is computed by the engine from Eqs.~\eqref{eq:affect} and~\eqref{eq:neuroticism} and supplied to the decision prompt together with its categorical level and currently active stimuli. The model returns the appraisal: \texttt{situation} indicates whether the pedestrian interprets its circumstances as \texttt{normal}, \texttt{unsure}, or \texttt{danger}, while \texttt{urgency} ranges from 0 to 1 and indicates how strongly the agent judges that immediate protective action is needed.

The appraisal channel draws on representations acquired during model pretraining. For each prompt, including the symbolic patch, persona text, affective context, and retained memory, the model returns the label that best matches the description within its learned representation space. The available categories (\texttt{danger}, \texttt{unsure}, and \texttt{normal}) and their graded urgency are therefore derived from the model's learned statistical regularities. A large language model (LLM) maps a textual description of the current emergency scenario into a high-dimensional semantic representation and establishes associations between the current input and previously learned semantic patterns within its parameterized knowledge and representation space. Through the Transformer architecture, attention mechanisms, contextual modeling, and statistical regularities encoded in its parameters, the LLM performs contextual inference over the available situational information and generates the most probable and contextually appropriate interpretation of the current emergency scenario.

This capability represents an important distinction between large language models and conventional rule-based models in emergency and evacuation applications. Rule-based models primarily rely on explicitly predefined conditions, rules, thresholds, and decision pathways, whereas LLMs can leverage contextual information and learned semantic representations to generalize and reason about scenarios that have not been explicitly specified in the predefined rule set. In the context of occupant evacuation and emergency response, this enables the model to integrate heterogeneous situational descriptions and infer context-dependent interpretations that may not be directly encoded as explicit evacuation rules. Consequently, the appraisal is also model-dependent: another model may interpret the same symbolic patch differently. 

\subsubsection{Time Step, Movement, and Contention}\label{sec:2.3.6}

The simulation advances in discrete physical steps. Each active entity moves by at most one cell per step, either orthogonally (0.4~m) or diagonally (0.57~m), so the speed ceiling is 1.2~m\,s$^{-1}$ along the grid axes and 1.7~m\,s$^{-1}$ diagonally. Each pedestrian advances along its accepted, model-selected route and waits after exhausting its legal steps. Because each decision occupies a fixed interval of simulated time (Section~\ref{sec:2.3.1}), this time scale supports direct comparison of decision timing, route choice, and responses to neighboring behavior \citep{chens2017,vantollw2021}.

Each grid cell can hold one entity, providing an explicit spatial constraint and a well-defined procedure for resolving contention. Consistent with grid-based models that resolve simultaneous choices through cell occupancy and local contention \citep{zhuk2020,kirchnera2003}, when several entities select the same cell, one is chosen at random and the others are assigned feasible cells within their Moore neighborhoods. Opposing swaps, diagonal crossings, and static occupancy are resolved in the same pass, and diagonal movement cannot cut between two blocked orthogonal cells. Because resolution occurs during execution, the agents' stated preferences remain unchanged \citep{chenl2018}. Agents still active at the time limit are recorded as unresolved, distinguishing them from those reached by the adversary.

\subsubsection{Algorithmic Adversary and Evidence Propagation}\label{sec:2.3.7}

The adversary uses the adult physical model: one cell per turn, a 12~m sight radius, a 120\textdegree{} cone plus Moore neighborhood, and the common collision and terrain rules. Its distinct physical capability is defined by a 2.0~m attack radius. A fixed algorithmic policy keeps hazard behavior constant across pedestrian conditions.

The default policy is a perception-limited patrol--pursuit strategy. When no pedestrian is visible, the adversary patrols evacuation exits 1--3 and entrance gates 1--2 in sequence before returning to \texttt{exit1}. When a pedestrian enters its perception map, the adversary selects the nearest visible pedestrian, locks onto that target for six turns, and updates the last observed cell while visual contact persists. After losing sight, it searches toward the last observed cell for up to eight turns. If it does not reacquire a pedestrian, it resumes the patrol from its current patrol index. Routes are generated by the same physical-map A* procedure used for other algorithmic movement. This perception-limited policy resembles patrol-then-pursuit formulations used in campus-attack simulations \citep{campusattack2022}.

Threat information propagates through local observation and person-to-person communication. The prompt requires situation=danger but leaves the action to the model, with \texttt{move}, \texttt{wait}, and \texttt{turn\_and\_observe} all permitted. The recorded \texttt{situation} therefore measures prompt compliance, whereas the recorded action is a behavioral outcome. The engine preserves the returned fields and limits movement during the pending interval to the pedestrian's existing legal route. After the source leaves view, the prior direct observation remains part of the pedestrian's knowledge. An anonymous shout supports \texttt{unsure} but contains no threat coordinate.

Shouting carries the signal beyond the line of sight. A pedestrian who has decided to escape, either after a direct sighting or through the model's situation assessment, shouts while moving. Listeners within 15~m receive the message even without line of sight, and the message still conveys the bearing and approximate distance of the shouter, which also tell the one relative direction of danger. They evaluate it together with other available evidence when selecting an intention.

The shared policy defines the local-map rules and tool contract while leaving situation assessment, intention, and route choice to the pedestrian. An uninformed pedestrian may tour, observe, wait, or explore a legal route or target. Direct evidence and communicated information enter the next model observation, after which the model determines whether to adopt evacuation as its objective. Personality, memory, affect, and evidence can influence preferences among legal choices but cannot change glyph meanings, coordinates, or passability. Fig.~\ref{fig:framework} summarizes this methodological boundary, and Fig.~\ref{fig:perception} details the construction of each private spatial observation.

\section{Experimental Design}\label{sec:3}

This section applies the general framework in Section~\ref{sec:2} to an evacuation scenario in Plaza. The plaza is represented by a $155\times164$ raster with a cell length of 0.4~m, giving a raster extent of approximately $62.0\times65.6$~m and a walkable area of 1{,}201.12~m\textsuperscript{2}. Twenty-five groups of trees form fixed internal obstacles. Eleven adult pedestrians are distributed across the plaza, while one moving threat enters from a fixed location. Five openings connect the plaza to its surroundings: two 4.0-m-wide entrance gates at the northern and southern ends and 3 evacuation exits that never closed along the eastern boundary, with widths of 1.4, 3.5, and 1.1~m.

The entrance gates and evacuation exits have different operational functions. During normal use, the northern and southern gates admit people into the plaza. Once the simulated attack activates the emergency phase, both entrance gates begin closing and subsequently become impassable. This rule represents emergency access control: admitting additional people after a violent incident would expose them to the threat and increase the number of potential victims. The three designated evacuation exits are therefore opened and remain available for outward movement. The asymmetric opening rule follows the functional distinction in the actual plaza between stopping further entry and maintaining evacuation capacity.

All simulations use the same plaza geometry, opening-control rule, pedestrian number, adult perception parameters, movement rules, threat policy, communication radius, memory mechanisms, and language model. The experimental factor is the personality composition of the eleven-person group. Each profile uses the fixed T-scores listed in Table~\ref{tab:persona}. The analysis examines whether personality composition is associated with situation appraisal, urgency and affect, changes in objectives and routes, access to exit information, and the resulting evacuation outcomes under the same physical and informational constraints.

The experiment evaluates eight personality compositions in eight matched blocks.The analysis combines run-level outcomes with pedestrian-level records of perception, decisions, memory, movement, information transfer, and final status. The following subsections specify the plaza and common parameters, outcome measures, statistical treatment, model configuration, and experimental conditions.

\subsection{Simulation Setup and Common Parameters}\label{sec:3.1}

\subsubsection{Spatial Layout and Emergency Opening Control}\label{sec:3.1.1}

\begin{figure*}[pos=tp]
\centering
\includegraphics[width=\textwidth]{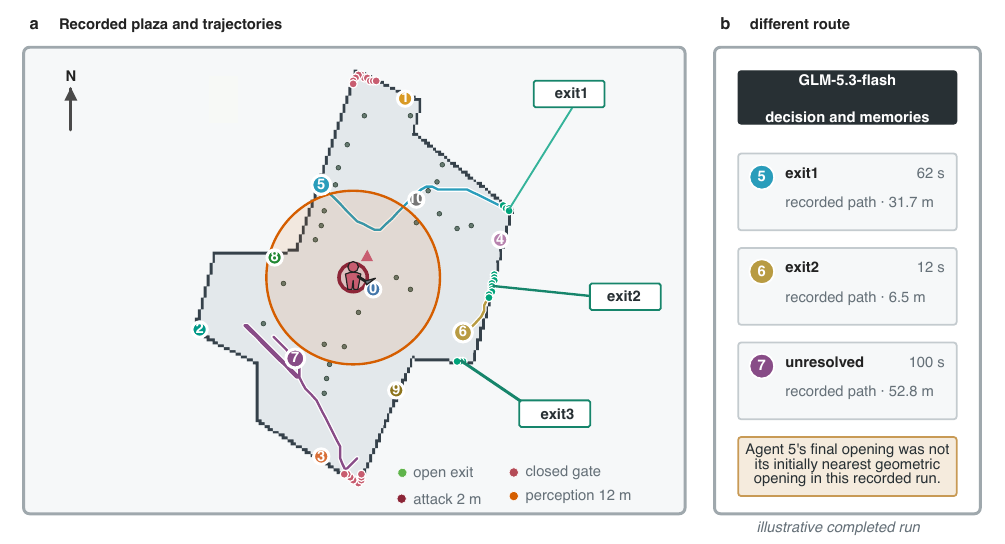}
\caption{\D{Plaza geometry and three illustrative trajectories. The perception map shows the three evacuation exits, two entrance gates, eleven pedestrian starting cells, fixed obstacles, and the initial threat position. The concentric circles indicate the implemented 2~m attack radius and 12~m perception radius. The trajectories illustrate three recorded outcomes from one completed run.}}
\label{fig:plaza-trajectories}
\end{figure*}

Fig.~\ref{fig:plaza-trajectories} shows the spatial relation among the two entrance gates, three evacuation exits, fixed obstacles, pedestrian starting locations, and the initial threat location. The entrance gates occupy the northern and southern ends of the plaza, whereas the three evacuation exits are distributed along the eastern boundary. The 25 groups of trees are distributed within the walkable interior and divide the open plaza into locally constrained movement corridors.

The state of each opening forms part of the time-dependent physical grid $G^t$. During the emergency phase, an entrance passes through open, closing, and closed states; once closed, it is no longer physically passable. The three evacuation exits remain open in the reference scenario. Pedestrians do not receive these states from the complete physical map. An agent learns that an opening is available or unavailable only by observing it directly or by receiving information from another pedestrian. An entrance observed before closure may therefore remain in the agent's memory as apparently usable until later evidence updates that belief. The opening-control rule consequently affects both physical route availability and the information on which agents base subsequent decisions.

For the Plaza scenario, each simulation contains eleven adult pedestrians placed at spatially distributed starting cells, together with one moving threat-"terrorist" entering from a fixed cell. Within each matched block, all personality-composition conditions use the same pedestrian starting cells, threat starting cell, plaza geometry, obstacle arrangement, and opening-control schedule. Personality profiles are reassigned among the fixed pedestrian starting cells across blocks so that the results are not tied to a particular profile occupying a particular location.

Each physical step lasts $\Delta t=1/3$~s and permits movement to one neighbouring cell. With a cell length of 0.4~m, the maximum speed is 1.2~m\,s$^{-1}$ along the grid axes and approximately 1.7~m\,s$^{-1}$ diagonally. When several agents attempt to enter the same cell, one is selected at random and the others are assigned feasible neighbouring cells. Movement prevents overlap, opposing swaps, diagonal crossings, and passage between blocked corners. A simulation ends when every pedestrian has evacuated or been reached by the threat, or when 240 simulated seconds have elapsed. Pedestrians still active at the time limit are recorded as unresolved.

\subsubsection{\D{Outcome Measures}}\label{sec:3.1.2}

The outcome measures describe evacuation outcomes, threat exposure, and pedestrian behavior.

\begin{table}[pos=htbp]
\caption{\D{Outcome measures computed for each pedestrian.}}
\label{tab:metrics}
\centering
\small
\setlength{\tabcolsep}{4pt}
\renewcommand{\arraystretch}{1.14}
\begin{tabular}{@{}>{\raggedright\arraybackslash}p{.36\linewidth}>{\raggedright\arraybackslash}p{.56\linewidth}@{}}
\toprule
\textbf{Measure} & \textbf{Definition} \\
\midrule
\multicolumn{2}{@{}l}{\textit{Evacuation}}\\
Evacuation time & Turns from run start to reaching an opening \\
Trajectory length & Distance walked along the executed trajectory \\
Evacuation efficiency & Shortest exit distance over walked distance, capped at 1; zero when unresolved \\
Critical safety margin & Smallest distance to the adversary over the run \\
Mean threat distance & Mean distance to the adversary over the run \\
Outcome & Evacuated, reached by the threat, or unresolved at the time limit \\
\midrule
\multicolumn{2}{@{}l}{\textit{Behavioral descriptors}}\\
Pre-movement delay & Turns from first evidence to the first committed move \\
Movement directness & Net displacement divided by path length over the 10~s following a decision, ending at evacuation, death, or the simulation limit \\
Reversals per 100~m & Number of heading changes beyond 120\textdegree{}, normalized per 100~m of walked distance \\
Herding rate & Fraction of moves aligned within 45\textdegree{} with neighbors inside 5~m \\
Pocket cells entered & Entries into cells with at most two passable orthogonal neighbors \\
Exit used, familiar-exit use & Which opening was taken, and whether it was the nearest known one \\
Threat response & Situation appraisal and action after directly seeing \texttt{A} or \texttt{D} \\
\bottomrule
\end{tabular}
\end{table}

\subsubsection{\D{Statistical Treatment}}\label{sec:3.1.3}

Movement directness was calculated as net displacement divided by path length over the 10~s following each applied decision, ending at evacuation, death, or the simulation limit and including the final movement step. The analysis included 10{,}947 decisions after initialization from 704 pedestrians. Urgency and directness were divided at their pooled medians; each pedestrian's share of decisions in a quadrant was then used to compare outcomes and calculate AUC.

The 5 homogeneous compositions were compared using run-level decision shares within the 8 paired blocks. Two-sided paired sign-flip tests enumerated all $2^8$ assignments, with Holm adjustment across the 10 pairwise comparisons for each measure. The 95\% intervals in Fig.~\ref{fig:arousal-efficacy} were obtained by resampling the paired blocks.

Logistic regression compared the 462 evacuated and 202 killed pedestrians, while linear regression of log evacuation time included the evacuated pedestrians. Geometry was represented by initial distance to the nearest designated evacuation exit; personality by composition; information by exit-information categories, the numbers of distinct warning senders and recipients, and whether and when threat evidence was first recorded; and affect by mean reported urgency, mean engine-computed fear, and the share of high-urgency, low-directness decisions. Each factor group was fitted separately and then included in a combined model. McFadden pseudo-$R^2$ describes the evacuation-outcome models, and ordinary $R^2$ describes the time models. These models summarize associations within the recorded simulations.

\subsubsection{\D{Language Model Configuration}}\label{sec:3.1.4}

The decision and memory roles both used \texttt{GLM-5.3-flash}, with a sampling temperature of 0.2. Ordinary decisions used the model's lowest reasoning level, and decisions triggered by direct visual evidence used a high level.

\subsection{\D{Experimental Conditions}}\label{sec:3.2}

The executed experiment varied only the personality composition of the eleven-person group. Eight compositions were defined using the five profile templates in Section~\ref{sec:2.1.1}. Five were homogeneous: \textbf{A} comprised eleven average pedestrians, \textbf{B} eleven calm-independent pedestrians, \textbf{C} eleven bold-solitary pedestrians, \textbf{D} eleven rule-following pedestrians, and \textbf{H} eleven anxious-conformist pedestrians. Conditions D--H formed a dose ladder in which rule-following pedestrians were progressively replaced by anxious-conformist pedestrians: \textbf{E} contained eight rule-following and three anxious-conformist pedestrians, \textbf{F} contained six and five, and \textbf{G} contained three and eight, respectively. The run-level anxious-conformist proportions were therefore $0$, $3/11$, $5/11$, $8/11$, and $1$ across D--H. The five homogeneous conditions A, B, C, D, and H were compared pairwise within blocks. We conducted a post hoc spatial analysis of decisions based on remembered threat information. For each threat-avoidance decision with a recorded last sighting and a nonzero first executed step, we measured changes in distance to the last observed position and to the adversary's actual position at step onset. Both reference positions were held fixed when calculating these changes. Directional mismatch denoted a step that increased the former distance but decreased the latter. Mismatch rates were calculated among steps moving away from the last observed position, with 95\% percentile intervals obtained by resampling the 8 paired blocks. Evidence age was the time since last sighting reported in the decision input.

\section{\D{Results}}\label{sec:4}
\subsection{\D{Decision, Memory, and Evacuation Outcomes}}\label{sec:4.1}

This section examines how the recorded behaviors emerged, and to what extend the knowledge graph works.Of the 704 pedestrians, 462 (65.6\%) evacuated, 202 (28.7\%) were killed. The following analyses examine how pedestrians revised their decisions, how responses differed across personality compositions, and how exit information was associated with evacuation outcomes.

\subsubsection{\D{Decision Revisions during Evacuation}}\label{sec:4.1.1}

The knowledge graph illustrates each single decision-turn in the whole simulation. Across 10{,}947 decisions following each pedestrian's initial decision, plan-level revisions occurred in 64.5\% of epochs, compared with 8.8\% for goal-level revisions and 9.9\% for execution-level adjustments. Changes at different layers could occur within the same epoch, and it also means the pedestrian would switch exit indeed, and all of these are determined by the large language model.

\begin{figure*}[pos=tp]
\centering
\includegraphics[width=\linewidth]{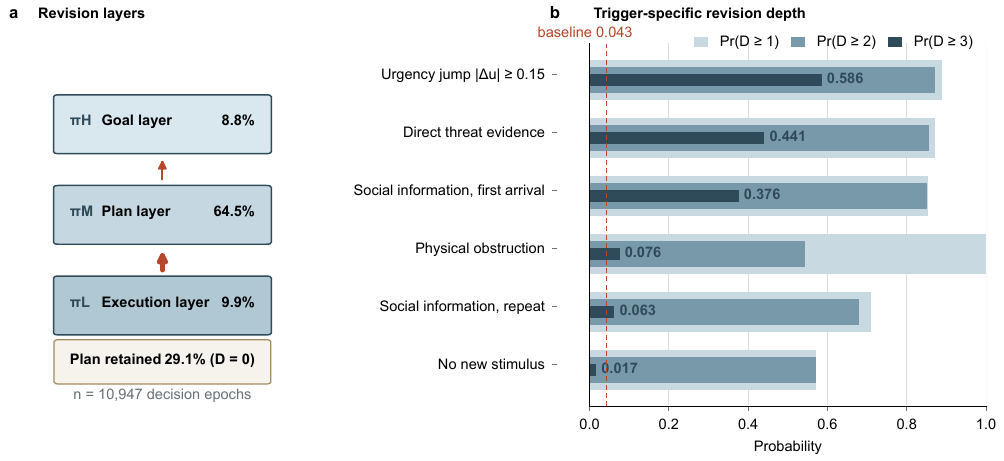}
\caption{\D{Decision revision by layer and stimulus. Left: the frequencies of goal-, plan-, and execution-level revisions across 10{,}947 decisions, together with the proportion retaining the previous decision. Right: cumulative revision probabilities for six stimulus classes. The dashed line marks the 4.3\% goal-revision rate for epochs with an absolute urgency change below 0.15.}}
\label{fig:revision-layers}
\end{figure*}

Goal revisions were most frequent when urgency changed substantially (Fig.~\ref{fig:revision-layers}). An absolute urgency change of at least 0.15 was accompanied by a goal revision in 58.6\% of epochs, compared with 4.3\% when the change was smaller. New direct threat evidence was accompanied by a goal revision in 44.1\% of epochs. The association between urgency changes and goal revision persisted across thresholds of 0.10, 0.15, and 0.20, with corresponding risk ratios ranging from 12.27 to 14.37.

Information from other pedestrians was associated with more goal revisions on first receipt than on subsequent receipt: 37.6\% versus 6.3\%. Physical constraints were accompanied by revisions at the plan level or above in 54.3\% of decisions and goal-level revisions in 7.6\%. Pedestrians therefore usually retained their goals while adjusting their plans or local movement in response to physical constraints.

Once pedestrians left ordinary activity, none returned to it. Threat avoidance and movement toward a known exit were retained at the next decision with probabilities of 0.982 and 0.931, respectively. These patterns indicate that evacuation-related objectives were largely maintained while plans and local movement continued to change.

\subsubsection{\D{Personality, Appraisal, and Arousal}}\label{sec:4.1.2}

Danger appraisal differed most across personality profiles when pedestrians relied on remembered threat information. Without threat information, danger-assessment rates were nearly identical across the 5 homogeneous personality compositions, differing by at most 0.002. When pedestrians relied on remembered threat information, the gap widened to 26.5 percentage points, with the danger-assessment rate in the anxious-conformist composition reaching 52.8\%, approximately twice the 26.3\% observed in the calm-independent composition. With direct sightings, the gap in danger-assessment rates narrowed to 3.3 percentage points across the 5 personality compositions. These results show that personality-related differences in danger appraisal depend on the available threat information, with remembered evidence revealing the clearest separation between profiles.

We examined the relationship between reported urgency and subsequent movement directness. Across the evaluated decisions, 32.5\% combined high urgency with high movement directness, making this the largest of the 4 quadrants. The mean share of high-urgency, low-directness decisions was 31.6\% among pedestrians who were killed, approximately 4.5 times the 7.0\% among those who evacuated, with an AUC of 0.839 for distinguishing the 2 outcomes.  The opposite combination showed a clear contrast: low-urgency, high-directness decisions accounted for an average of 62.9\% among evacuated pedestrians and 15.0\% among killed pedestrians, with an AUC of 0.802. Evacuated pedestrians also had the lowest mean urgency and highest mean directness, with a centroid at $(u=0.358,\ \hat d=0.867)$, compared with $(0.653,\ 0.630)$ for killed pedestrians and $(0.587,\ 0.416)$ for unresolved pedestrians. These results show that reported urgency gains behavioral meaning when considered alongside subsequent movement: pedestrians who were killed more often combined a strong sense of urgency with less direct movement, revealing a gap between the urgency expressed by the agent and the movement that followed.

Across the 5 homogeneous personality compositions, the mean share of high-urgency, low-directness decisions ranged from 11.41\% for calm-independent agents to 34.33\% for rule-following agents, a roughly 3-fold difference. Anxious-conformist agents followed at 33.32\%, while average and bold-solitary agents recorded 29.45\% and 22.52\%, respectively. Calm-independent agents had lower shares than both rule-following and anxious-conformist agents in all 8 paired blocks (Holm-adjusted $p=0.078$ for each comparison).  For low-urgency, high-directness decisions, calm-independent agents recorded the highest mean share at 31.08\%, approximately 2.4 times the 12.72\% recorded by anxious-conformist agents. The calm-independent composition therefore combined the lowest share of high-urgency, low-directness decisions with the highest share of the opposite combination.

\begin{figure*}[pos=tp]
\centering
\includegraphics[width=\linewidth]{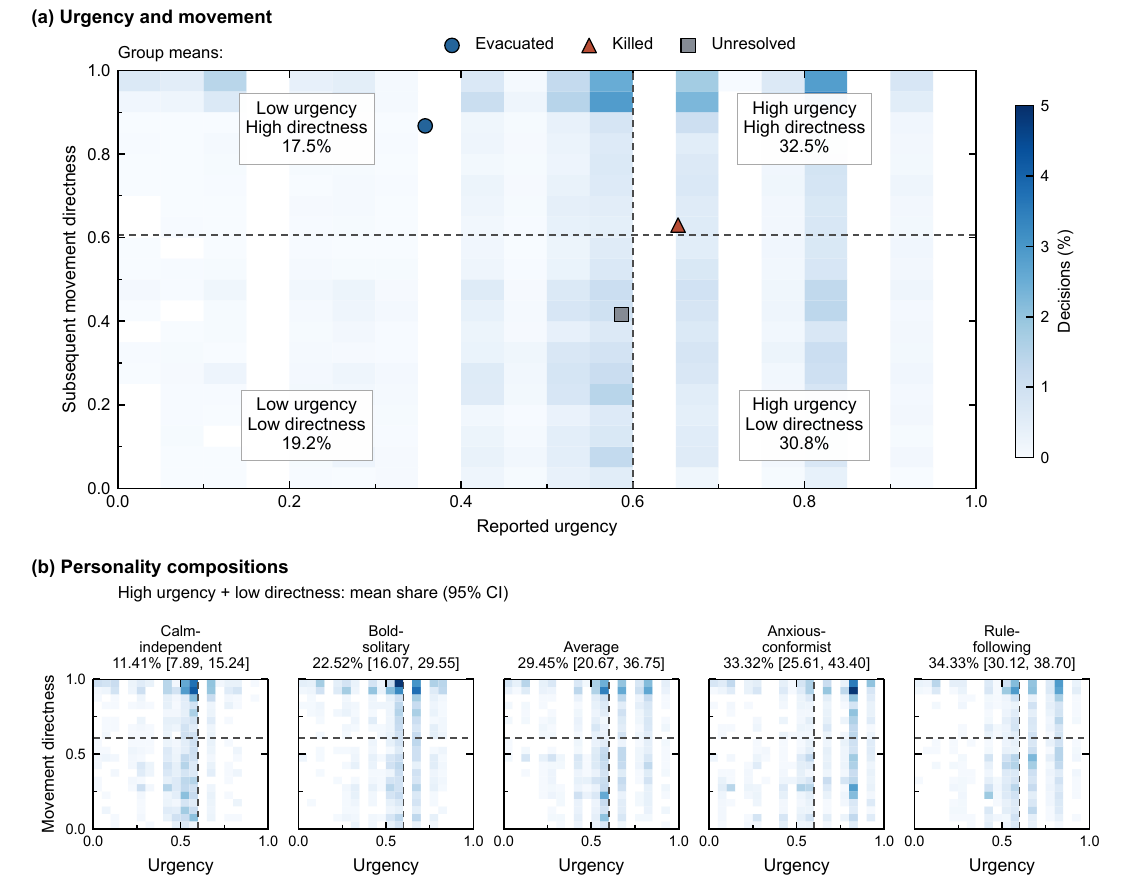}
\caption{\D{Reported urgency and subsequent movement directness across 10{,}947 decisions. (a) Dashed lines mark the pooled medians of urgency ($u=0.60$) and directness ($\hat d=0.606$); quadrant labels give the percentages of all evaluated decisions. Symbols mark outcome-group means, with each pedestrian weighted equally. (b) The 5 homogeneous personality compositions. Values above each panel give the mean run-level share of high-urgency, low-directness decisions and its 95\% confidence interval from paired-block resampling. In all panels, the linear color scale shows the percentage of decisions within that panel falling in each cell. The individual share of high-urgency, low-directness decisions distinguished death from evacuation with an AUC of 0.839.}}
\label{fig:arousal-efficacy}
\end{figure*}

In conclusion, the outcome and personality comparisons identify reported urgency and subsequent movement directness as complementary descriptors of simulated evacuation behavior (Fig.~\ref{fig:arousal-efficacy}).

Across 102{,}045 memory operations involving 34{,}489 distinct items, writing accounted for the largest share at 33.8\%, followed by forgetting at 29.9\% and rewriting at 24.7\%, while explicit retention made up the remaining 11.6\%. Anxious-conformist agents stood out for retaining existing memories: their explicit-retention share reached 15.6\%, compared with 10.8--12.1\% across the other 4 homogeneous personality compositions. They also recorded the highest share of decisions without revision, at 33.5\%, showing that more frequent memory retention accompanied greater decision continuity in this composition. 
Across the knowledge graphs, each decision was linked to an average of 5.51 memory items. Fig.~\ref{fig:knowledge-graph} shows these links for one pedestrian. Table~\ref{tab:memory-transitions} traces 2 cases in which remembered observations, their interpretation, and escape plans changed over successive decisions.

\textit{In Case 1, earlier shouts acquired a new meaning after the pedestrian saw an armed person.} The pedestrian initially wanted to observe and clarify the situation; after the sighting, its memory described the earlier shouts as probable warnings and placed escape ahead of exploration. When the known gate subsequently closed, the recorded plan combined continued threat avoidance with a search for an open exit.

This sequence is consistent with the distinction between interpreting warning cues and assessing protective actions in the Protective Action Decision Model \citep{lindell2012padm}. The memory revisions documented both why the pedestrian intended to leave and which escape options it considered usable.

\textit{In Case 2, recognizing that a gate was closed preceded abandoning the route toward it.} The pedestrian recorded the closure but retained a plan to approach the gate and check whether it could be opened, and the next decision continued toward it. A later decision treated the route as impassable and sought an alternative, which the following memory revision recorded.

Experiments have shown that exit familiarity influences evacuation choices \citep{kinateder2018exit}. In the present case, the stated intention to verify whether a known gate could be used offers a possible explanation for continuing toward it after learning that it was closed. The record identifies an intermediate stage in route revision: the pedestrian acknowledged the obstruction while still considering whether the original escape plan could work.

\begin{table*}[pos=tp]
\caption{Changes in remembered information and escape plans in 2 illustrative cases.}
\label{tab:memory-transitions}
\centering
\small
\setlength{\tabcolsep}{4pt}
\renewcommand{\arraystretch}{1.15}
\begin{tabularx}{\linewidth}{@{}>{\raggedright\arraybackslash}p{0.12\linewidth}>{\raggedright\arraybackslash}p{0.37\linewidth}>{\raggedright\arraybackslash}p{0.20\linewidth}>{\raggedright\arraybackslash}X@{}}
\toprule
\textbf{Stage} &
\textbf{Memory excerpt (translated)} &
\textbf{Interpretation} &
\textbf{Decision or plan} \\
\midrule

\multicolumn{4}{@{}l}{\textbf{Case 1: From uncertain signals to escape and exit search}} \\
\addlinespace[3pt]

1. Situation unclear &
``I wanted to stop and observe to clarify the situation, without hastily following the others.'' &
The meaning of others' behavior is still unclear. &
The stated plan is to observe before deciding whether to follow others. \\
\addlinespace[5pt]

2. Threat confirmed &
``The 2 shouts I heard earlier \ldots\ were probably warnings.''
``Escaping comes first; exploration can wait.'' &
Earlier sounds are now interpreted as warnings. &
The decision changes from continuing activity to moving away from the threat. \\
\addlinespace[5pt]

3. Another exit needed &
``Next, I need to find an exit that is still open or a direction with more people, and keep moving away from the northeast.'' &
Escape remains necessary, but the known gate may be unusable. &
The next decision maintains threat avoidance and includes looking for other exits. \\
\midrule

\multicolumn{4}{@{}l}{\textbf{Case 2: From choosing a known exit to checking and abandoning it}} \\
\addlinespace[3pt]

1. Known exit chosen &
``I remembered that [the gate] was to the south, so I decided to go south, away from the armed person and toward the known exit.'' &
The known exit provides a destination for escape. &
The selected goal is to reach the known exit. \\
\addlinespace[5pt]

2. Closed gate to be checked &
``The exit door to the south is closed \ldots\ Once I get closer, I may check whether it can be opened or look for another route.'' &
The gate is closed, but the pedestrian still considers whether it can be used. &
The next decision continues toward the gate to check it. \\
\addlinespace[5pt]

3. Original route abandoned &
``The south exit gate is still closed and cannot be used, so I have switched to going east \ldots\ to look for another exit.'' &
The gate is now regarded as unusable. &
The decision selects an alternative route, and the revised memory records the new plan. \\
\bottomrule
\end{tabularx}

\par\smallskip
\begin{minipage}{\linewidth}
\footnotesize
Excerpts are translated from model-generated Chinese memories. Ellipses mark omissions; square brackets replace an internal gate identifier with a descriptive name. Stages are ordered within each case, and the final column summarizes the corresponding decision or stated plan. Both cases involve bold-solitary pedestrians in separate runs.
\end{minipage}
\end{table*}
An illustrative comparison involved 2 pedestrians who started from the same position in one paired block. Following warning shouts, the calm-independent pedestrian's memory emphasized moving south while watching for signs and continuing to search for an exit. In contrast, the anxious-conformist pedestrian's memory emphasized following the crowd along a familiar southern route. At the next recorded decision, their objectives were exit exploration and following others, respectively, while both continued moving southward. The memory--decision links thus distinguish different behavioral intentions accompanying a broadly similar movement direction, showing how the reconstructed graphs support interpretation of individual evacuation trajectories.

\subsubsection{\D{Information Access and Survival}}\label{sec:4.1.3}

Across all 704 pedestrians, knowledge of a usable exit was strongly associated with outcome. Pedestrians who acquired knowledge of at least one usable exit reached safety in 89.5\% of cases; among those who acquired no such knowledge, 1.05\% reached safety and 80.0\% were killed. Fisher's exact test yielded an odds ratio of 800.7 ($p=6.1\times10^{-118}$). The relationship was graded: rescue rates were 1.05\% for pedestrians who knew no usable exit, 64.3\% for those who knew one, and 95.7\% for those who knew two. At the run level, the number of pedestrians with usable-exit knowledge correlated with the number rescued ($\rho=0.607$, $p=1.0\times10^{-7}$).

\begin{figure*}[pos=tp]
\centering
\includegraphics[width=\linewidth]{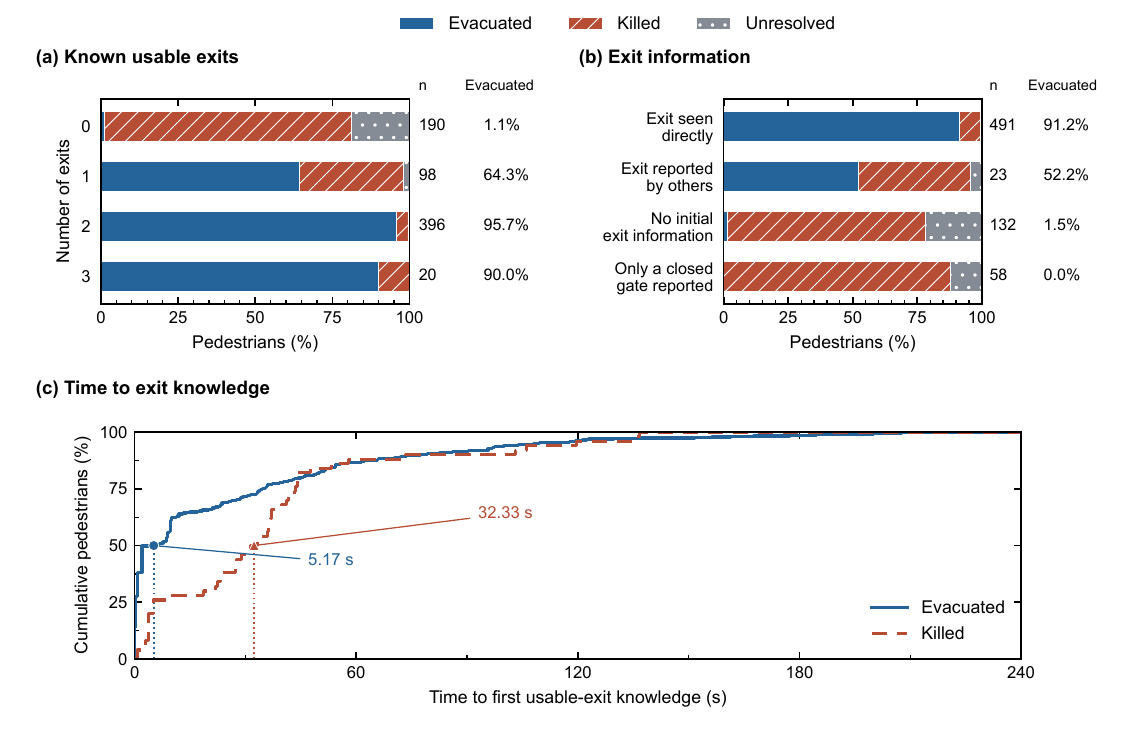}
\caption{\D{Exit knowledge and evacuation outcomes. (a) Outcome shares by the number of usable exits known. (b) Outcome shares for the 4 exit-information groups. ``No initial exit information'' refers to the absence of system-provided exit information at initialization; ``Only a closed gate reported'' refers to information received from others during the simulation. Numbers beside the bars give group sizes ($n$) and evacuation rates. (c) Cumulative distributions of time to first usable-exit knowledge among pedestrians who acquired it, grouped by final outcome. Markers indicate median times of 5.17~s for evacuated pedestrians and 32.33~s for killed pedestrians.}}
\label{fig:exit-knowledge}
\end{figure*}

The value of the information also depended on its content. Pedestrians who directly observed a usable exit were rescued in 91.2\% of cases, compared with 52.2\% of those informed by a peer. None of the 58 pedestrians whose only exit information referred to a closed gate was rescued, and 87.9\% were killed. Acquisition timing also differed by outcome: the median time to usable-exit knowledge was 5.17~s for rescued pedestrians and 32.33~s for killed pedestrians. Fig.~\ref{fig:exit-knowledge} compares evacuation outcomes by exit knowledge and information group, together with the timing of knowledge acquisition. Panel (a) shows evacuation rates of 1.05\% with no usable exit known, 64.3\% with 1, and 95.7\% with 2. Panel (b) includes the 58 pedestrians whose only reported exit information concerned a closed gate; none evacuated. Panel (c) shows median acquisition times of 5.17~s for evacuated pedestrians and 32.33~s for killed pedestrians.

The two propagation channels were associated with different outcomes. The number of distinct pedestrians reached by a shout yielded an AUC of 0.808 for being killed ($p=6\times10^{-42}$), while the number of distinct shouters heard yielded an AUC of 0.791 ($p=2\times10^{-34}$). By contrast, usable-exit knowledge predicted rescue ($p=7\times10^{-98}$). In this framework, shouts primarily carried alarm, whereas exit messages conveyed routes.  Among 311 pedestrians whose previous appraisal was normal and whose first evaluated decision containing a shout had no direct or remembered threat evidence, 286 (92.0\%) reported higher urgency than in their preceding decision. At that point, 254 (81.7\%) assessed the situation as unsure, and only 1 assessed it as dangerous. These results show that heightened urgency and uncertainty about danger can coexist in the simulated response to a warning.

Exit information propagated through a narrow channel. Each run had an average of 2.1 senders, and the two most active senders accounted for 90.1\% of receiving relations. Propagation extended no more than one hop: 591 pedestrians observed an exit directly, while 64 learned of one from someone who had observed it. Fig.~\ref{fig:info-networks} contrasts the two channels within a single run. The same eleven pedestrians occupy identical positions and are encoded by outcome in both panels, so only the edges differ. Shouting forms a dense, multi-source network, whereas exit information travels through a small number of directed, single-hop links.

\begin{figure*}[pos=tp]
\centering
\includegraphics[width=\linewidth]{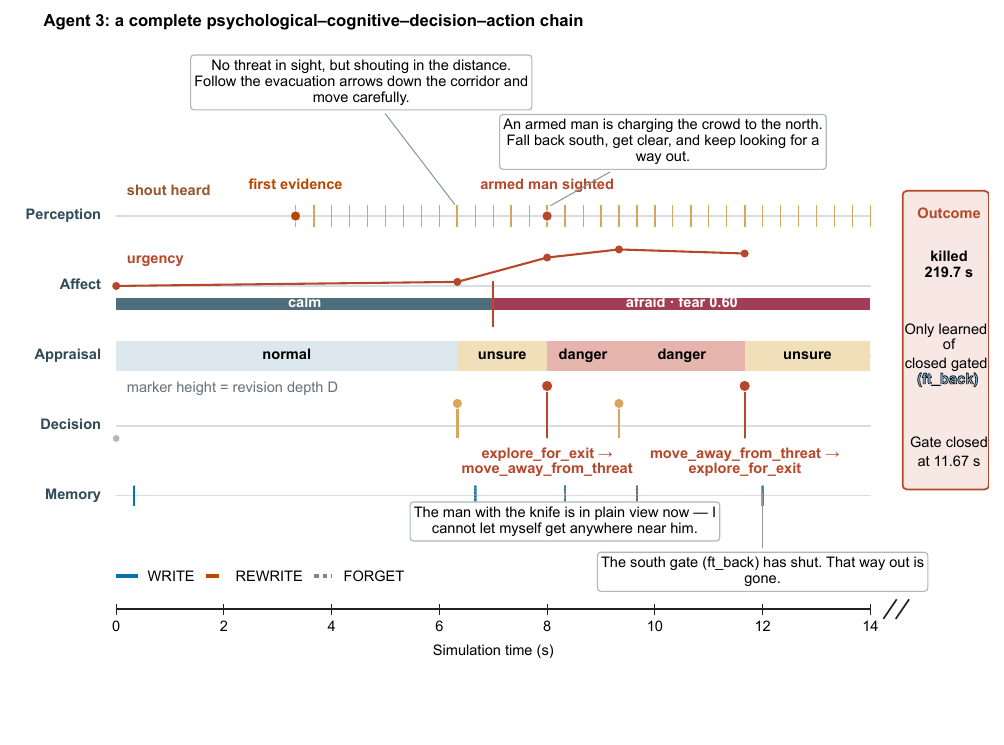}
\caption{\D{The recorded chain for Agent~3 in the rule-following run. Five lanes carry perception events, the affect transition and the self-reported urgency trace, the appraisal band, each decision at a height set by its revision depth, and the write, rewrite, and forget operations on memory. Callouts reproduce selected memory entries. The pedestrian learns of only one opening, the south entrance gate, which closes at 11.67~s.}}
\label{fig:evolution-chain}
\end{figure*}

Fig.~\ref{fig:evolution-chain} traces this process for one pedestrian. Five aligned lanes show perception events, the affect transition and urgency trajectory, the appraisal category, decisions positioned by revision depth, and memory write, rewrite, and forget operations. The two goal-layer revisions appear as the tallest decision markers; callouts reproduce the memory entries recorded around them, and a broken time axis connects the early sequence to the outcome at 219.7~s. Agent~3 in the rule-following condition of block~5 heard shouting at 3.33~s, changed from calm to afraid at 7.00~s, and observed the armed adversary at 8.00~s. The agent then changed its goal from exit seeking to threat avoidance and raised its self-reported urgency to 0.9 while moving toward the only opening it knew. When the gate closed at 11.67~s, the agent revised its goal again; it was killed at 219.7~s. Its opening knowledge referred only to the entrance gate that closed while it was approaching; meanwhile, entrance gates were closing with merely one person being able to go out of them.

Exit information was closely associated with whether pedestrians evacuated. The information-only model had a McFadden pseudo-$R^2$ of 0.6650, compared with 0.3541 for affect, 0.0537 for geometry, and 0.0036 for personality composition; the combined model reached 0.7020. Exit knowledge and communication records therefore captured much of the observed separation between evacuation and death.

\subsubsection{\D{Factors Associated with Evacuation Time}}\label{sec:4.1.4}

Evacuation time was associated with geometry, information, and affect. Their separate models had $R^2$ values of 0.6608, 0.6864, and 0.6859, respectively, compared with 0.0031 for personality composition; the combined model reached 0.8738. The higher fit of the combined model indicates that these factors describe evacuation duration more fully when considered together.
\section{{Discussion}}\label{sec:5}
\subsection{\D{Contribution to Evacuation Modeling}}\label{sec:5.1}

The framework connects the development of individual interpretations to the movement of a crowd. Social-force and cellular-automaton models describe local movement and interaction, while behavioral extensions represent differences in perception, preferences, and route choice. The present approach adds an explicit record of how an agent describes its situation, retains experience, and revises a plan within a shared physical environment.

Learning-based crowd models offer complementary ways to represent adaptive behavior. Deep reinforcement learning develops policies through environmental interaction \citep{zhangd2023,pangc2023}; hierarchical and knowledge-guided methods organize decisions at different levels \citep{sunl2026}; and attention-based planners coordinate movement in pedestrian environments \citep{dongl2024,lij2026}. The contribution of the present framework is to make the information available at a decision, the model's stated interpretation, and the resulting movement available for joint analysis.

This approach connects movement analysis with questions about how decisions develop. Similar trajectories can accompany different intentions, while a continuing goal can be pursued through changing routes. Recording these distinctions makes it possible to examine how an evacuation unfolds from the perspective of an individual with incomplete information.

\subsection{\D{Memory, Interpretation, and Route Revision}}\label{sec:memory-discussion}

Memory carries both an account of the situation and the conditions attached to an intended action. The cases in Table~\ref{tab:memory-transitions} distinguish 2 processes: new evidence changes the meaning assigned to earlier cues, and new knowledge about an exit is evaluated against an existing escape plan. These processes connect successive decisions through an evolving understanding of what has happened and what remains possible.

The distinction between recognizing a change and acting on its implications is particularly relevant to evacuation. A pedestrian may acknowledge a closed gate while retaining an intention to verify whether it can be used, leaving an intermediate stage between receiving information and abandoning a route. This interpretation extends the case analysis into a testable explanation of route persistence and is consistent with the separation of cue interpretation and protective-action assessment in the Protective Action Decision Model \citep{lindell2012padm}.

The reconstructed knowledge graph makes this sequence open to examination by linking retained information to the decisions at which it was available. It supports questions about which interpretations persist, which plans remain conditional, and what evidence accompanies their revision? The resulting analysis treats evacuation as a developing sequence of judgments and actions, with memory preserving the connections between them.

\subsection{\D{Scope and Limitations}}\label{sec:5.2}

The findings concern a small pedestrian population in one plaza under a specified moving-threat policy. Personality compositions were compared within a common environment, and the personality representation entered both the decision prompts and the affect model. The observed differences therefore concern the combined personality configuration.

The recorded interpretations are model-generated accounts supplied within the decision process. Their correspondence with human judgments, route choices, and responses to communicated information remains to be established. The present contribution is a framework and a set of observable decision sequences through which such correspondence can be investigated.

\subsection{\D{Implications for Future Evacuation Research}}\label{sec:future-directions}

A central opportunity is to evaluate evacuation models against the development of behavior as well as its final outcome. Human experiments could compare when participants recognize danger, how they interpret an uncertain exit, and when they abandon a preferred route. These comparisons would test whether simulated transitions reproduce the sequence and timing of human responses.

The same framework could support controlled studies of evacuation communication. Messages that report a closure could be compared with messages that explain its consequences and identify a usable alternative, testing whether information content changes verification, route revision, and exposure to danger. Such studies would connect the interpretation of a warning to the movement that follows it.

A further direction is to connect individual decision histories with collective evacuation patterns. Distributions of perceived danger, exit knowledge, and continuing intentions could inform behavioral states in mesoscopic models and help investigate changes in aggregate exit demand. This would extend the use of LLM-powered simulation from examining individual choices to studying how uneven information and different interpretations develop into crowd-level behavior.

\section{\D{Conclusions}}\label{sec:6}

This study develops an LLM-powered framework for examining how evacuation decisions evolve under a moving threat. Private perception, personality, and retained experience provide the context for individual choices, while the physical simulation records how those choices are executed. Linking memory revisions to decisions and movement makes the formation, continuation, and revision of escape plans accessible to analysis.

Most decision changes occurred at the plan level. Large changes in urgency and new direct threat evidence were associated with goal revisions, while physical constraints usually prompted adjustments that preserved the current goal. Once pedestrians left ordinary activity, they maintained evacuation-related objectives while adapting their plans and movement.

Personality compositions differed in danger appraisal under remembered evidence and in the relationship between reported urgency and movement. The share of high-urgency, low-directness decisions ranged from 11.41\% for calm-independent agents to 34.33\% for rule-following agents. The calm-independent composition maintained lower shares than the rule-following and anxious-conformist compositions across all paired blocks.

Knowledge of a usable exit was strongly associated with evacuation. Pedestrians with knowledge of a usable exit were rescued in 89.5\% of cases, compared with 1.05\% of those without such knowledge, yielding an odds ratio of 800.7. No pedestrian whose only exit information identified a closed gate was rescued. Exit knowledge traveled at most one hop, through an average of 2.1 senders per run; the two most active senders accounted for 90.1\% of transmissions. Rescued pedestrians acquired this knowledge at a median of 5.17~s.

Exit information was closely associated with whether evacuation occurred, while evacuation time was associated with geometry, information, and affect together. These findings connect the knowledge available to pedestrians with both the completion and duration of evacuation.

The memory cases identify changes in the meaning of earlier information and in the conditions for retaining an escape plan. These sequences provide concrete hypotheses about delayed responses and persistence toward a known exit. The framework opens a path toward evacuation research that connects how people understand a developing emergency with how their individual choices contribute to collective movement.

\clearpage
\section*{\D{Supplementary Figures}}
\setcounter{figure}{0}
\renewcommand{\thefigure}{S\arabic{figure}}

\begin{figure}[pos=htbp]
\centering
\includegraphics[width=\linewidth]{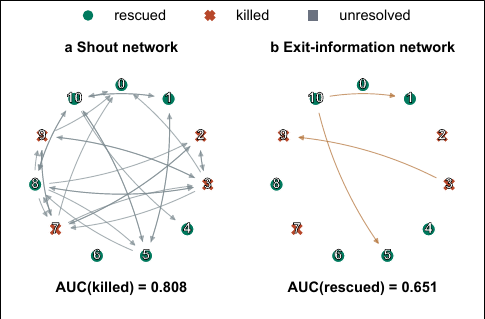}
\caption{\D{Shout and exit-information propagation in one run. The same eleven pedestrians hold identical positions in both panels and are encoded by outcome. Directed edges give deduplicated sender--receiver pairs for (a) heard shouts and (b) transmitted exit information. Across the experiment, shout out-degree carries an AUC of 0.808 for being killed and exit knowledge an AUC of 0.651 for rescue.}}
\label{fig:info-networks}
\end{figure}

\begin{figure}[pos=htbp]
\centering
\includegraphics[width=\linewidth]{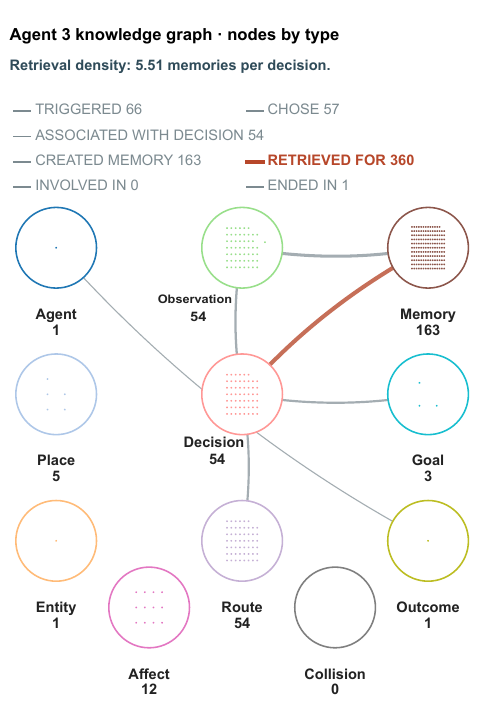}
\caption{\D{Reconstructed knowledge graph for Agent~3. Nodes are grouped into eleven type clusters and relations are bundled between types, with \texttt{RETRIEVED FOR} emphasized. This agent's graph holds 54 observations, 54 decisions, 54 routes, 163 memories, and 12 affect records, with an average of 6.67 memory links per decision. Across the experiment, the graphs contain 75{,}944 nodes and 138{,}963 edges.}}
\label{fig:knowledge-graph}
\end{figure}

\section*{\D{Acknowledgements}}

This work was supported by the National Key Research and Development Program of China (No. 2025YFB2606402), the National Science Foundation of China (No. 72604081), and the Sichuan Provincial Science and Technology Innovation Project (No. 2024YFHZ0345).

\FloatBarrier
\bibliographystyle{cas-model2-names}
\bibliography{references}
\end{document}